\documentclass[%
aip,
 amsmath,amssymb,
 reprint,%
]{revtex4-1}

\usepackage{graphicx}
\usepackage{dcolumn}
\usepackage{bm}

\usepackage[utf8]{inputenc}
\usepackage[T1]{fontenc}
\usepackage{mathptmx}
\usepackage{xcolor}

\begin{document}

\renewcommand\thesection{\arabic{section}}
\renewcommand{\thesubsection}{\thesection.\arabic{subsection}}

\title[Sample title]{Machine learning surrogate models for particle insertions and element substitutions}

\author{Ryosuke Jinnouchi}
\affiliation{ 
Toyota Central R\&D Labs., Inc., 41-1 Yokomichi, Nagakute, Aichi, 480-1192, Japan 
}%
 \email{jryosuke@mosk.tytlabs.co.jp}

\date{\today}

\begin{abstract}
Two machine learning-aided thermodynamic integration schemes to compute the chemical potentials of atoms and molecules have been developed and compared. One is the particle insertion method, and the other combines particle insertion with element substitution. In the former method, the species is gradually inserted into the liquid, and its chemical potential is computed. In the latter method, after the particle insertion, the inserted species is substituted with another species, and the chemical potential of this new species is computed. In both methods, the thermodynamic integrations are conducted using machine-learned potentials trained on first-principles datasets. The errors of the machine-learned surrogate models are further corrected by performing thermodynamic integrations from the machine-learned potentials to the first-principles potentials, accurately providing the first-principles chemical potentials. These two methods are applied to compute the real potentials of proton, alkali metal cations, and halide anions in water. The applications indicate that these two entirely different thermodynamic pathways yield identical real potentials within statistical error bars, demonstrating that both methods provide reproducible real potentials. The computed real potentials and solvation structures are also in good agreement with past experiments and simulations. These results indicate that machine learning surrogate models enabling the atomic insertion and element substitution provide a precise method for determining the chemical potentials of atoms and molecules.
\end{abstract}

\maketitle

\section{\label{section1} Introduction}

The chemical potential of atoms and molecules in condensed matter is a crucial property that determines many physical characteristics, including the coexistence points of different phases, concentrations of minority species, solubility in solvents, free energy changes of chemical reactions, and redox potentials of electrochemical reactions. One of the ultimate goals of molecular dynamics (MD) simulations is to predict this property. However, accurate prediction using first principles (FP) methods is a highly challenging task. The difficulty can be well understood by considering the calculation of the actual potential of one or more atoms in a liquid. The real potential is defined as the change in free energy when the solute is transferred from a vacuum just outside the liquid surface into the liquid. The simplest approach to compute this free energy change is to perform thermodynamic integration (TI)~\cite{Kirkwood_JCP_1935, Zwanzig_JCP_1954} using the particle insertion method, where the interactions between the solute and the liquid are gradually switched on.~\cite{Kelvin_JCP_2009, Duignan_CS_2017, Dorner_PRL_2018} While this brute-force method is simple, it suffers from poor statistical accuracy. Particularly in the initial stages, where an infinitesimally small interaction between the inserted solute and the liquid must be used, the integrand in TI can diverge to infinity because the not-yet-interacting atom can come very close to a solvent atom, experiencing a huge repulsive potential. Although this issue can be partially circumvented through variable transformations,~\cite{Dorner_PRL_2018} most FP codes become unstable when two atoms are very close. Moreover, beyond the initial steps, a dramatic change in the solvation structure along the coupling constant requires extensive sampling, posing a significant challenge for FP calculations.

Recently, machine learning force fields (MLFFs) have emerged as a powerful method that significantly accelerates the computation of free energy changes.~\cite{Morawietz_PNAS_2016, Grabowski_npjComputMater_2019, Jinnouchi_PRB_2020, Jinnouchi_JCP_2021, Jinnouchi_npjComputMater_2024, Jinnouchi_arXiv_2024} When trained with a sufficient number of training datasets, MLFFs can accurately reproduce the potential energy landscape.~\cite{Behler_PRL_2007, Bartok_PRL_2010, Thompson_JCP_2015, Shapeev_MMS_2016, Zhang_PRL_2018, Jinnouchi_PRB_2019, Drautz_PRB_2019, Batzner_NatCommn_2022, Ilyes_NEURIPS_2022} This enables MD simulations in TI calculations, which require statistical sampling, to be accelerated by several orders of magnitude. Although MLFFs can introduce non-negligible errors in free energy changes, these errors can be corrected by performing TI or TPT calculations from MLFFs to potential energies calculated using FP methods. Because the main portion of the free energy change is already evaluated by the MLFFs, the small residual deviations between the MLFF and FP potential energies can be corrected with just tens of picoseconds of MD simulations. To date, this ML-aided FP calculation scheme has been successfully applied in predicting the melting points of water~\cite{Morawietz_PNAS_2016} and inorganic materials,~\cite{Jinnouchi_PRB_2019} the anharmonic free energies of multielement alloys,~\cite{Grabowski_npjComputMater_2019} the reversible potentials of redox species,~\cite{Jinnouchi_npjComputMater_2024} and the real potentials of ions and adsorbates.~\cite{Jinnouchi_PRB_2020, Jinnouchi_JCP_2021, Jinnouchi_arXiv_2024}

\begin{figure*}
\centering
\includegraphics[width=0.80\textwidth,angle=0]{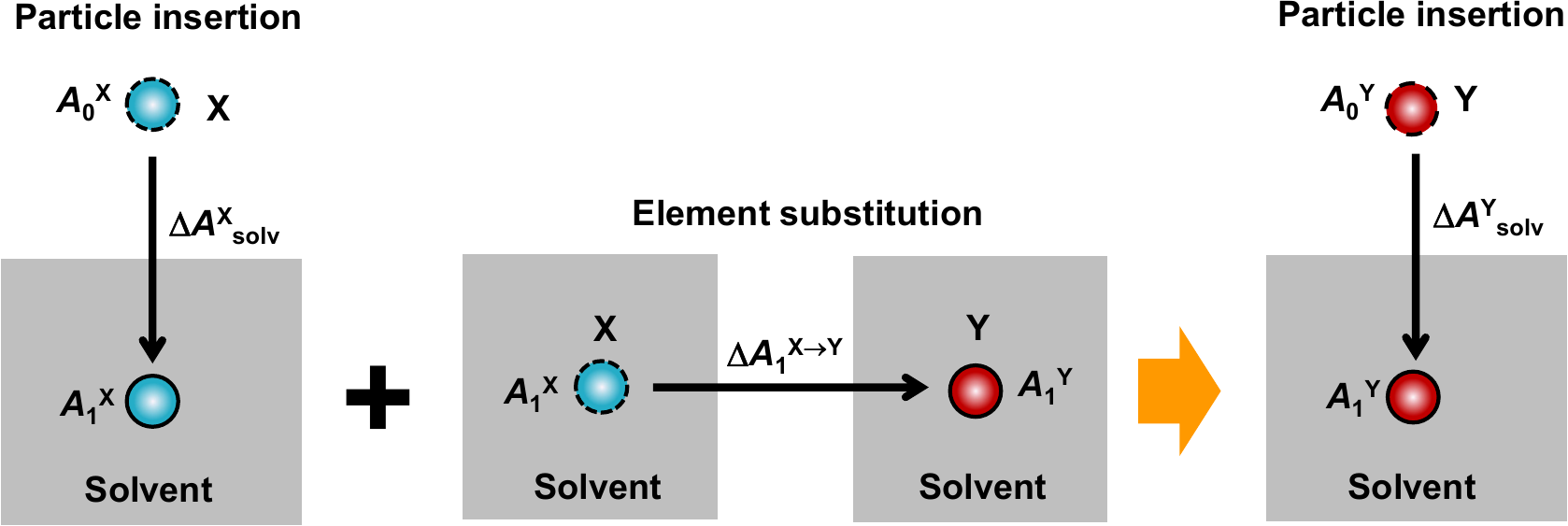}
\caption{Schematic of particle insertion and element substitution to compute the absolute solvation free energy.}
\label{fig1}
\end{figure*}

Despite significant acceleration, several challenges remain in ML-aided free energy computation. This method requires careful selection of the coupling path that smoothly connects two endpoints, similar to conventional TI calculations using empirical force fields.~\cite{Straatsma_JCP_1988, Simonson_MS_1993, Zacharias_JCP_1994, Grochola_JCP_2004} If the coupling path is irreversible, it may yield significant errors in the resulting free energy change. This issue is particularly crucial when performing TI calculations with the particle insertion method using MLFFs, as the system can become irreversibly trapped in an erroneously predicted stationary potential well if the MLFF is not trained on that stationary point. To prevent this problem, an improved ML-aided TI scheme was proposed.~\cite{Jinnouchi_JCP_2021, Jinnouchi_arXiv_2024} In this approach, an intermediate model potential is introduced to keep the system within the phase space covered by the training data. The coupling path is divided into two parts. The first integration proceeds from the non-interacting system to the intermediate model potential, and the second integration proceeds from the model potential to the MLFF. This method has successfully predicted the real potential of protons in water and hydroxides formed at the water-Pt(111) interface.~\cite{Jinnouchi_JCP_2021, Jinnouchi_arXiv_2024} However, performing TI of atomic insertion between two structurally distinct states, namely the non-interacting and interacting systems, using MLFFs presents a fundamental challenge. It requires sampling over a long thermodynamic pathway between two points with significantly different phase spaces. The issue of sampling speed is resolved by MLFFs, and as long as the model potentials can generate smooth thermodynamic paths within the phase space interpolated by the training data, this method can provide accurate free energy changes. However, constructing model potentials that can keep the system within the interpolatable phase space occasionally requires trial and error. Moreover, as with any free energy calculation, there are always concerns about the accuracy and reproducibility of results obtained from a single thermodynamic pathway, necessitating proper verification.

The main goals of the present work are two-fold. First, I aim to provide an alternative, smooth, and efficient thermodynamic pathway to verify the free energy changes computed by the particle insertion method. The proposed alternative pathway is based on substituting one element with another, as shown in Fig.~\ref{fig1}. When the chemical properties of the two elements are similar, it becomes much easier to construct a smooth coupling path between these two endpoints, as demonstrated in past TI calculations using conventional physics-based force fields.~\cite{Lybrand_JACS_1985, Lybrand_PNAS_1986, Straatsma_JCP_1988} Once the chemical potential of a chemical species is obtained through the atomic insertion method, the chemical potentials of other species can be successively determined using a simpler elemental substitution method in a domino-like fashion. Second, I aim to provide FP results for the real potentials of a proton, six alkali metal cations (Li$^+$, Na$^+$, K$^+$, Rb$^+$, Cs$^+$, Fr$^+$), and five halide anions (F$^-$, Cl$^-$, Br$^-$, I$^-$, At$^-$). Among these ions, Fr$^{+}$ and At$^{-}$ are radioactive, making their solutions hypothetical liquids. The solvation of alkali cations and halide anions has long been a subject of study using MD simulations due to its significance in biochemistry and physical chemistry.~\cite{Straatsma_JCP_1988, Dang_JCP_1993, Lee_JPC_1996, Toth_JCP_1996, Koneshan_JPCB_1998, Ramaniah_JCP_1999, White_JCP_2000, Lyubartsev_JCP_2001, Raugei_JACS_2001, Loeffler_JCP_2002, Hribar_JACS_2002, Spangberg_JPCB_2003, Ayala_JCP_2003, Grossfield_JACS_2003, Tongraar_PCCP_2003, Grossfield_JCP_2004, Heuft_F_JCP_2005, Heuft_I_JCP_2005, Lamoureux_JPCB_2006, Kelvin_JCP_2009, Dangelo_IC_2010, Rowley_JCTC_2012, Bankura_JCP_2013, Duignan_CS_2017, Ambrosio_JPCL_2018, Jinnouchi_PRB_2020} Because of the substantial computational cost, calculations of real potentials or solvation free energies using FP methods have only been performed for a few types of ions.~\cite{Kelvin_JCP_2009, Duignan_CS_2017, Ambrosio_JPCL_2018} To my knowledge, there are no systematic studies that have calculated the real potentials of all the aforementioned ion species using the TI employing the same FP method. In this work, I aim to demonstrate the solvation properties that FP calculations provide for these ions and to examine whether they are consistent with past experiments and previous computational results.

\section{\label{section2}Method}

\subsection{\label{section2_1}Particle insertion}

The real potential of species X is defined as the free energy difference between X in the gas phase just outside the water surface and X in the water. The real potential is the negative value of the work function of species X and differs from the solvation free energy by the product of the water surface potential and the charge of species X. As in the traditional definition, we define the concentration at standard states as 0.1 MPa (1/24.46 mol L$^{-1}$) for the gas phase and 1 mol L$^{-1}$ for the liquid phase. As explained later, actual TI computations were conducted using the same periodic cell (with a concentration of 0.87 mol L$^{-1}$) for both gaseous and aqueous species. The equations presented in this section are also derived for this condition. After the computations, the change in free energy caused by the difference in concentration was corrected using the ideal gas model.

In the present article, I propose two TI schemes shown in Fig.~\ref{fig1}: particle insertion and element substitution. This subsection explains the particle insertion method. The real potential is calculated as an integral of the expectation value of the derivative of the Hamiltonian with respect to the coupling parameter $\lambda$, which specifies a path from the reference state ($\lambda$ = 0) to the interacting system of interest ($\lambda$=1):~\cite{Kirkwood_JCP_1935, Zwanzig_JCP_1954}
\begin{align}
\alpha &= A_{1} - A_{0} = \int_{0}^{1} \left< \frac{\partial H^\mathrm{in}}{\partial \lambda} \right>_{\lambda} d\lambda. \label{eq1}
\end{align}
The reference state is set to be the non-interacting species X and the solvent without the species X. Along the thermodynamic pathway, the interaction between X and the solvent is gradually turned on. This transition can be realized using the Hamiltonian defined as follows:~\cite{Blumberger_JCP_2006, Dorner_PRL_2018}
\begin{align}
H^\mathrm{in} &= \sum\limits_{i=1}^{N_\mathbf{a}} \frac{|\mathbf{p}_{i}|^2}{2 m_{i}} + \lambda \left( U_\mathrm{1} + z \Delta \phi \right) + (1-\lambda) U_\mathrm{0}, \label{eq2}
\end{align}
where $\lambda$ is a coupling constant, which specifies a path from the species in vacuum and liquid water ($\lambda=0$) to the mixture of them ($\lambda=1$), $U _\mathrm{0}$ and $U_\mathrm{1}$ denote potential energies of these two endpoints, $z$ denotes the ionic charge. In any calculation using periodic boundary conditions, the energy changes due to the removal or addition of charged species are not entirely well-defined. To correct for this, the potential gap $\Delta \phi$ is introduced. This potential gap essentially specifies the potential of the vacuum level just outside a water surface, which serves as the common reference point in electrochemistry. The potential gap is determined through a separate slab calculation involving an interface between water and the vacuum. ~\cite{Cheng_PCCP_2012, Ambrosio_JCP_2015, Ambrosio_JPCL_2018, Jinnouchi_npjComputMater_2024, Jinnouchi_arXiv_2024}

As done in previous studies,~\cite{Jinnouchi_PRB_2020, Jinnouchi_JCP_2021, Jinnouchi_npjComputMater_2024, Jinnouchi_arXiv_2024} I decompose the TI equation [Eq.(\ref{eq1})] into two integrals:
\begin{align}
\alpha &= \Delta A_{\mathrm{I}} +\Delta A_{\mathrm{II}},  \label{eq3}  \\
\Delta A_{\mathrm{I}} &= \int_{0}^{1} \left< \frac{\partial H^\mathrm{in}_{\mathrm{I}}}{\partial \lambda_{\mathrm{I}}} \right>_{\lambda_{\mathrm{I}}} d\lambda_{\mathrm{I}},  \label{eq4}  \\
\Delta A_{\mathrm{II}} &= \int_{0}^{1} \left< \frac{\partial H^\mathrm{in}_{\mathrm{II}}}{\partial \lambda_{\mathrm{II}}} \right>_{\lambda_{\mathrm{II}}} d\lambda_{\mathrm{II}}.  \label{eq5}
\end{align}
In the first Hamiltonian $H_\mathrm{I}$ [Eq.~(\ref{eq4})], as in the previous studies,~\cite{Timothy_CS_2017} a Gaussian soft repulsive potential $U_\mathrm{model}$ is gradually added to the potential energy of the solvent without species X in order to create a space for species X:
\begin{align}
H^\mathrm{in}_{\mathrm{I}} &= \sum\limits_{i=1}^{N_{\mathrm{a}}} \frac{ \left| \mathbf{p}_{i} \right|^{2}}{2 m_{i}} + \lambda_{\mathrm{I}} U_\mathrm{model} + \sum\limits_{i \notin \mathrm{X}} U_{i} \left( 0 \right).  \label{eq6}
\end{align}
Here, $N_\mathrm{a}$ is the number of atoms in the system, and $U_{i} \left( 0 \right)$ is the atomic potential energy represented by a MLFF of the solvent without species X. In the second Hamiltonian, the solute-solvent interaction potential $U_\mathrm{model}$ is gradually replaced by the interaction potential described by an MLFF:
\begin{align}
H^\mathrm{in}_{\mathrm{II}} &= \sum\limits_{i=1}^{N_{\mathrm{a}}} \frac{ \left| \mathbf{p}_{i} \right|^{2}}{2 m_{i}} + \lambda_{\mathrm{II}} \left[ \sum\limits_{i=1}^{N_{\mathrm{a}}} U_{i}(1) + z \Delta \phi \right] +  \nonumber \\
&\left(1 - \lambda_{\mathrm{II}} \right) \left[ U_\mathrm{model} + \sum\limits_{i \notin \mathrm{X}} U_{i}\left( 0 \right) \right],  \label{eq7}
\end{align}
where $U_{i}(1)$ denotes the atomic potential energy of the solution system involving the species X. 

As in previous studies,~\cite{Bartok_PRL_2010, Jinnouchi_PRL_2019, Jinnouchi_PRB_2019} the atomic potential energies [$U_{i}\left( 0 \right)$ and $U_{i}\left( 1 \right)$] are represented as a linear combination of functions $K \left( \mathbf{x}_{i} \left( \kappa \right), \mathbf{x}_{i_\mathrm{B}} \right)$:
\begin{align}
U_{i}\left( \kappa \right) &= \sum\limits_{i_\mathrm{B}=1}^{N_\mathrm{B}} w_{i_\mathrm{B}} K\left(\mathbf{x}_{i}\left( \kappa \right) ,\mathbf{x}_{i_\mathrm{B}} \right). \label{eq8}
\end{align}
Here, the descriptors $\mathbf{x}_{i}\left( \kappa \right) = \mathbf{x}_{i} \left[ \rho_{i} \left( \mathbf{r}, \kappa \right) \right]$ are the functional of the density distribution function around atom $i$,
\begin{align}
\rho_{i}\left(\mathbf{r}, \kappa \right) &= \sum\limits_{j\notin \mathrm{X}} f_\mathrm{ cut} \left( \left| \mathbf{r}_{j} - \mathbf{r}_{i} \right| \right) g \left (\mathbf{r} - \left( \mathbf{r}_{j} - \mathbf{r}_{i} \right) \right) + \nonumber \\
&= \kappa \sum\limits_{j\in \mathrm{X}} f_\mathrm{ cut} \left( \left| \mathbf{r}_{j} - \mathbf{r}_{i} \right| \right) g \left (\mathbf{r} - \left( \mathbf{r}_{j} - \mathbf{r}_{i} \right) \right),  \label{eq9}
\end{align}
where the function $g$ is a smoothed $\delta$-function, and $f_\mathrm{cut}$ is a cutoff function that smoothly eliminates the contribution from atoms outside a given cutoff radius $R_\mathrm{cut}$. Hence, $U_{i}\left( 0 \right)$ and $U_{i}\left( 1 \right)$ represent the local energies without and with the group X, respectively. Using these notations, $U_{0}$ and $U_{1}$ can be written as $\sum\limits_{i \notin \mathrm{X}} U_{i}\left( 0 \right)$ and $\sum\limits_{i=1}^{N_{\mathrm{a}}} U_{i}\left( 1 \right)$. It should be noted that the MLFF of the fully interacting system $U_{1}$, trained on the FP data of the periodic system, is also shifted by $-z \Delta \phi$. This shift is corrected in Eq.~(\ref{eq7}).

\subsection{\label{section2_2}Element substitution}

The ML-aided particle insertion allows for accurate computation of the real potential. However, it requires many numerical integration grids because the integrand changes steeply along the coupling path from the non-interacting state to the fully interacting state. It should also be noted that $U_\mathrm{model}$ needs to be designed so that the system remains within the phase space where the MLFF can interpolate. If the real potential $\alpha^\mathrm{X}$ of species X is already given by the atom insertion method, the free energy $A_{1}^\mathrm{Y}$ of species Y, which is generated by substituting elements in X with other elements, can be obtained by computing the free energy change $\Delta A_{1}^\mathrm{X \rightarrow Y}$ associated with the substitution of the elements in the fully interacting system as follows:
\begin{align}
A_{1}^\mathrm{Y} &= A_{0}^\mathrm{X} + \alpha^\mathrm{X}+ \Delta A_{1}^\mathrm{X \rightarrow Y}, \label{eq10}
\end{align}
where $A_{0}^\mathrm{X}$ is the free energy of X in vacuum, which can be easily obtained by the ideal gas model. The free energy change $\Delta A_{1}^\mathrm{X \rightarrow Y}$ can be calculated as:
\begin{align}
\Delta A_{1}^\mathrm{X \rightarrow Y} &= \int_{0}^{1} \left< \frac{\partial H^\mathrm{sb}}{\partial \xi} \right>_{\xi} d\xi. \label{eq11} \\
H^\mathrm{sb} &= \sum\limits_{i=1}^{N_\mathbf{a}} \frac{|\mathbf{p}_{i}|^2}{2 m_{i}} + \xi \left( U_\mathrm{1}^\mathrm{Y} + z^\mathrm{Y} \Delta \phi \right) + (1-\xi) \left ( U_\mathrm{1}^\mathrm{X} + z^\mathrm{X} \Delta \phi \right). \label{eq12}
\end{align}
Here, the TI changes the potential from $U_\mathrm{1}^\mathrm{X}+ z^\mathrm{X} \Delta \phi$ to $U_\mathrm{1}^\mathrm{Y} + z^\mathrm{Y} \Delta \phi$ while keeping the mass of the species fixed to that of species X. This computation is particularly feasible when the chemical characteristics of species Y are similar to those of species X. The real potential of species Y can be obtained as follows:
\begin{align}
\alpha^\mathrm{Y} &= A_{1}^\mathrm{Y} - A_{0}^\mathrm{\prime Y}, \label{eq13}
\end{align}
where $A_{0}^\mathrm{\prime Y}$ is the free energy of the species Y in vacuum {\em with the mass of the species X}. The value can be also easily obtained by the ideal gas model.

\subsection{\label{section2_3}Correction to free energy of MLFF}

Although both the ML-aided particle insertion and element substitution methods provide the free energy change, they may introduce non-negligible errors due to inaccuracies in the MLFF models. These errors can be corrected through a TI calculation that transitions from the MLFF to the potential obtained by the FP method:~\cite{Jinnouchi_npjComputMater_2024, Jinnouchi_arXiv_2024}
\begin{align}
\Delta A_{\kappa}^\mathrm{FP-ML} &= \int_{0}^{1} \left< \frac{\partial H_{\kappa}^\mathrm{FP-ML}}{\partial \eta} \right>_{\eta} d\eta, \label{eq14} \\
H_{\kappa}^\mathrm{FP-ML} &= \sum\limits_{i=1}^{N_{\mathrm{a}}}  \frac{ \left| \mathbf{p}_{i} \right|^{2}}{2 m_{i}} + \eta U_{\kappa}^\mathrm{FP} + \left( 1-\eta \right) U_{\kappa}^{\mathrm{ML}}, \label{eq15}
\end{align}
where $U_{\kappa}^\mathrm{ML}$ and $U_{\kappa}^\mathrm{FP}$ represent the potential energies for the state specified by $\kappa$ ($\kappa$=0 and $\kappa$=1 denote the non-interacting and fully interacting states, respectively). These energies are calculated using the MLFF and FP methods, respectively. Utilizing $\Delta A_{0}^\mathrm{FP-ML}$ and $\Delta A_{1}^\mathrm{FP-ML}$, the free energy change based on the FP method $\Delta A^\mathrm{FP}$ is calculated as:
\begin{align}
\Delta A^\mathrm{FP} &= \Delta A^\mathrm{ML} + \Delta A_{1}^\mathrm{FP-ML} - \Delta A_{0}^\mathrm{FP-ML}, \label{eq16}
\end{align}
where $\Delta A^\mathrm{ML}$ is the free energy change determined by the particle insertion or element substitution method described in the previous subsection.

The correction of MLFF errors through this TI is generally considered necessary to verify the accuracy of the results. However, in the present application, all MLFFs were highly accurate, and this correction resulted in only a very small change, altering the calculated real potentials by an average of 50 meV.

\subsection{\label{section2_4}MLFF generations}

\begin{figure}
\centering
\includegraphics[width=0.35\textwidth,angle=0]{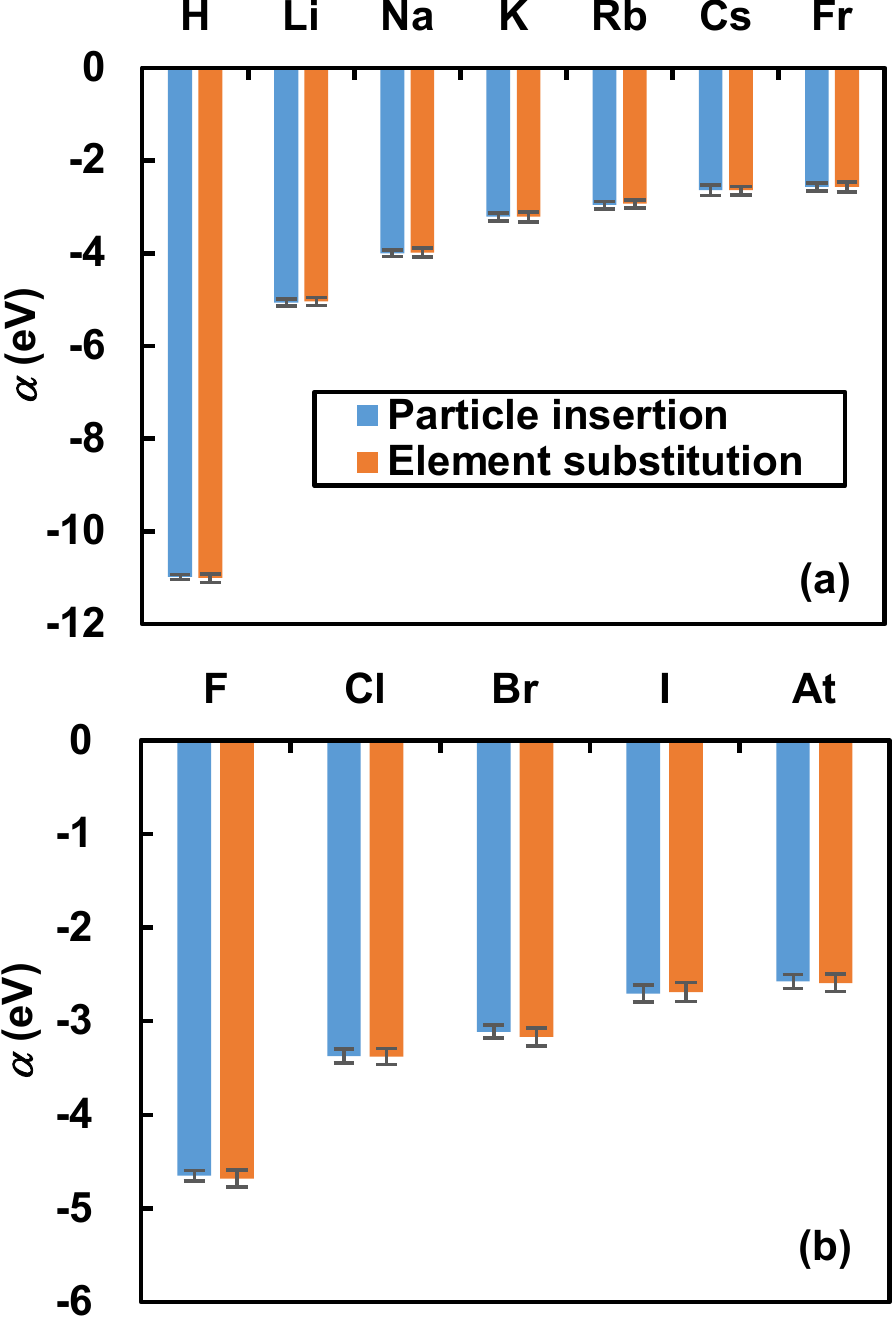}
\caption{Computed real potentials of alkali cations (a) and halide anions (b). Error bars represent 2$\sigma$ determined by block averaging analysis.}
\label{fig2}
\end{figure}

In both the training runs for generating the MLFFs and the production runs for computing free energy changes, the liquid solvent was modeled using a unit cell with a side length of 12.4 \AA, containing 64 water molecules. For the TI calculations, a single ion per unit cell was inserted, as illustrated in Fig. \ref{fig2}. The choice of the 64-water-molecule system was based on a previous study~\cite{Jinnouchi_npjComputMater_2024} that showed this system size achieved sufficient convergence

All calculations were carried out using the Vienna Ab initio Simulation Package (VASP).~\cite{Kresse_PRB_1996, Kresse_CMS_1996} For the MLFF models, I employed the kernel-based algorithm detailed in previous publications.~\cite{Jinnouchi_PRB_2019, Jinnouchi_JPCL_2020} Similar to the pioneering methods,~\cite{Behler_PRL_2007, Bartok_PRL_2010} the potential energy is expressed as a sum of atomic energies. Each atomic energy is represented as a weighted sum of kernel basis functions of the radial and angular descriptors.~\cite{Jinnouchi_JCP_2020} The weights on these kernel basis functions are optimized to best reproduce the FP training datasets. These optimizations are performed within a Bayesian framework, which allows for accurate predictions of energies, forces, stress tensor components, and their uncertainties, thereby enabling efficient on-the-fly sampling of reference structures during FPMD simulations of the target systems.

To calculate the solvation free energy for each ion using the particle insertion method, a single MLFF was generated. The parameters for the descriptors and kernel basis function optimized in the previous study~\cite{Jinnouchi_JPCL_2023} were utilized. Their values are listed in Supplementary Table~S1. The MLFF was trained on liquid water on-the-fly during a 100-ps NVT-ensemble heating MD simulation, where the temperature was increased from 300 K to 500 K. Subsequently, the MLFF was trained on-the-fly during another 100-ps NVT-ensemble heating MD simulation of the solution containing the target ion. The MLFF underwent further training on-the-fly during TI calculations along the coupling constant $\lambda_\mathrm{II}$, which seamlessly transitions the soft repulsive Gaussian potential to the full interaction potential via Eq. (\ref{eq7}). In this training run, a 6-point Gauss-Lobatto quadrature was employed. At each integration grid point, a 100-ps NVT-ensemble MD simulation was performed. A single MLFF for each ion was also generated using the element substitution method. The initial conditions before the on-the-fly training during the TI calculations were the same as those for the MLFFs for the particle insertion method. After these training runs, the MLFF was further trained during TI calculations along the coupling constant $\eta$. In this training run, a 5-point equispaced grid was used, and a 100-ps NVT-ensemble MD simulation was performed at each integration grid point. All MD simulations utilized the Nose-Hoover thermostat.~\cite{Nose_JCP_1984, Hoover_PRA_1985} During the heating runs, the temperature was linearly ramped up from 300 K to 500 K. For efficient sampling, the mass of hydrogen atoms was increased to 4.0 amu, and the timestep of the MD simulations was increased to 2.0 fs.

The FP calculations were performed using plane wave basis sets with a cutoff energy of 520 eV and the projector augmented wave (PAW) method.~\cite{Kresse_PRB_1999} The PAW atomic reference configurations were 1s$^{1}$ for H, 2s$^{2}$2p$^{4}$ for O, 2s$^{2}$2p$^{5}$ for F, 2p$^{6}$3s$^{1}$ for Na, 3s$^{2}$3p$^{5}$ for Cl, 3s$^{2}$3p$^{6}$4s$^{1}$ for K, 4s$^{2}$4p$^{5}$ for Br, 4s$^{2}$4p$^{6}$5s$^{1}$ for Rb, 5s$^{2}$5p$^{5}$ for I, 5s$^{2}$5p$^{6}$6s$^{1}$ for Cs, 6s$^{2}$6p$^{5}$ for At, and 6s$^{2}$6p$^{6}$7s$^{1}$ for Fr. The $\Gamma$-point was used for Brillouin zone integration. In all calculations, the exchange-correlation interactions between electrons were modeled using the revised Perdew-Burke-Ernzerhof semi-local functional (RPBE+D3),~\cite{Hammer_PRB_1999} augmented with Grimme's dispersion interaction.~\cite{Grimme_JCP_2010, Grimme_JCC_2011}

In all on-the-fly sampling, a threshold spilling factor of 0.01 was used. If the spilling factor~\cite{Miwa_PRB_2016, Jinnouchi_PRB_2019} exceeded 0.01, the MLFF prediction for the structure appearing in the MD simulation was considered to involve large errors. In such cases, an FP calculation was performed, the structure and the FP data were added to the training dataset, the MLFF was reconstructed, and the atomic positions were updated using the FP result. Otherwise, the MLFF prediction was used to update the atomic positions. This approach allowed us to bypass more than 99 \% of FP calculations, enabling highly efficient sampling.
Similar to the MLFFs in previous studies,~\cite{Jinnouchi_PRB_2019, Zuo_JPCA_2020, Jinnouchi_JPCL_2020, Jinnouchi_JCP_2020, Jinnouchi_JCP_2021, Verdi_npjComputMater_2021, Lysogorskiy_npjComputMater, Jinnouchi_JPCL_2023, Jinnouchi_npjComputMater_2024, Hijes_JCP_2024} the MLFFs achieve small RMSEs of 0.6 meV atom$^{-1}$ and 0.036 eV \AA$^{-1}$ (see error distributions in Supplementary Figs.~S1 to~S4, enabling accurate and efficient computations of free energy changes.

\subsection{\label{section2_5}Production runs}

\begin{figure*}
\centering
\includegraphics[width=0.90\textwidth,angle=0]{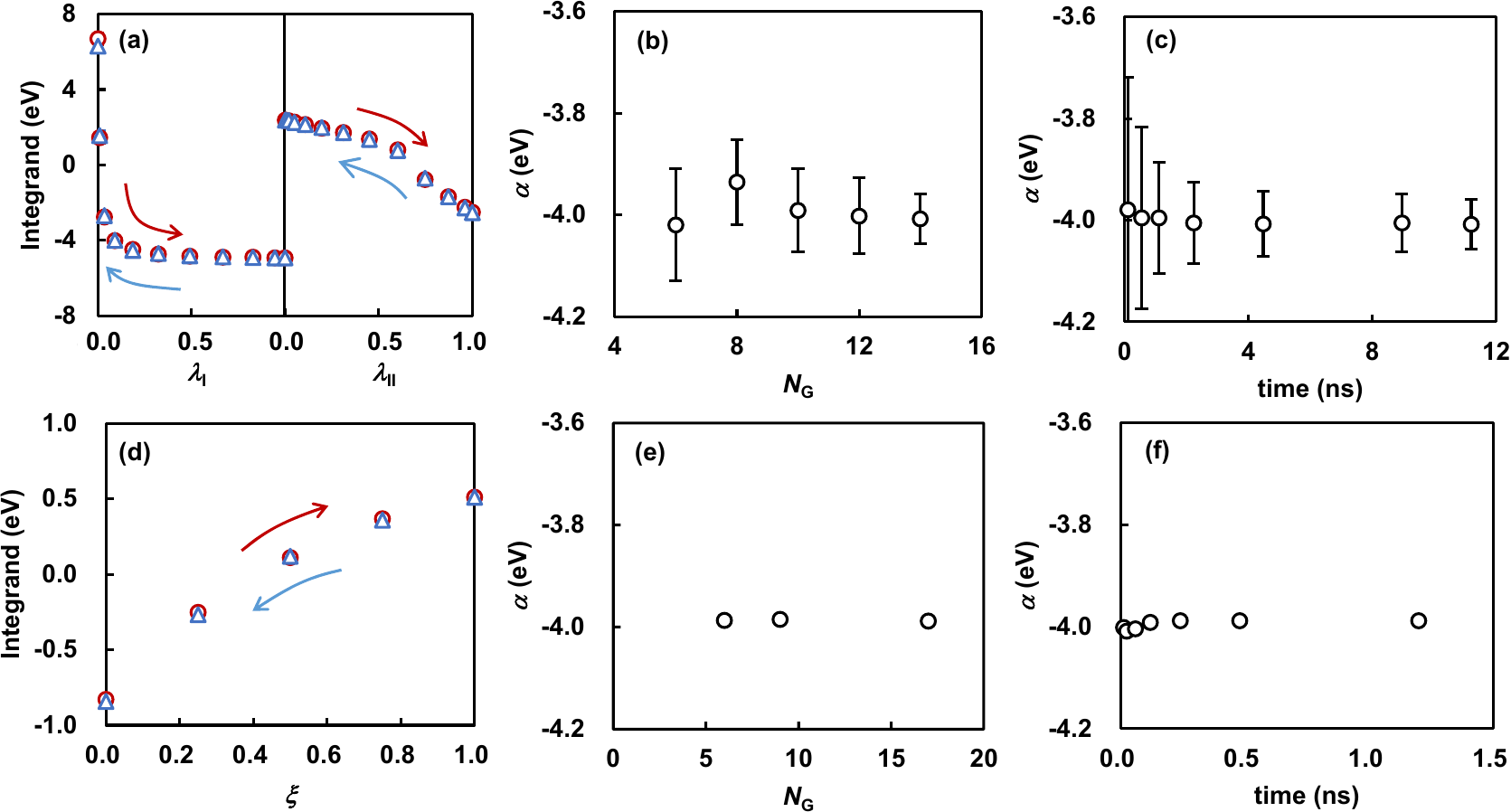}
\caption{Numerical results for the real potential of the Na$^{+}$ cation: (a) integrand of the TI calculation for particle insertion, (b) the real potential of the single Na$^{+}$ cation $\alpha$ computed by the particle insertion method as a function of the number of integration grids $N_\mathrm{G}$, and (c) $\alpha$ as a function of the total MD simulation time. Figures (d) to (f) show the same information for the element substitution from K$^{+}$ to Na$^{+}$, subsequently executed after the K$^{+}$ insertion. Error bars represent 2$\sigma$ determined by block averaging analysis. Error bars in (d) to (f), which are too small to be visible, indicate those generated solely by the element substitution.}
\label{fig3}
\end{figure*}

The TI calculations by the particle insertion method were conducted using the variable transformation proposed by Dorner and co-workers:~\cite{Dorner_PRL_2018}
\begin{align}
\lambda_\mathrm{I,II} &= \left( \frac{x + 1}{2} \right)^{\frac{1}{1-k}}. \label{eq17}
\end{align}
The variable $k$ was set to 0.5 in this study. After verifying the convergence with respect to the number of numerical integration grids [see examples in Fig.~\ref{fig3} (b)], the 12-point and 14-point Gauss-Lobatto quadratures were employed for the coupling constants $\lambda_\mathrm{I}$ and $\lambda_\mathrm{II}$, respectively. As in previous publications, the TI calculations were conducted in both forward and reverse directions along the coupling parameters to check the reversibility of the thermodynamic path. For the proton, which diffuses via the complex Grotthuss mechanism, 14 cycles of forward and reverse transitions were performed to achieve good statistics. For other ions, the number of cycles was set to 4-6.

The TI calculations using the element substitution method were conducted for the following pairs: H$^{+}\leftrightarrow$Li$^{+}$, Li$^{+}\leftrightarrow$Na$^{+}$, Na$^{+}\leftrightarrow$K$^{+}$, K$^{+}\leftrightarrow$Rb$^{+}$, Rb$^{+}\leftrightarrow$Cs$^{+}$, Cs$^{+}\leftrightarrow$Fr$^{+}$, F$^{-}\leftrightarrow$Cl$^{-}$, Cl$^{-}\leftrightarrow$Br$^{-}$, Br$^{-}\leftrightarrow$I$^{-}$, and I$^{-}\leftrightarrow$At$^{-}$. Similar to Eq.(\ref{eq17}), the variable transformation from $\xi$ to $x$ was employed for the element substitution of H$^{+}$$\leftrightarrow$Li$^{+}$, where the integrand changed steeply near the endpoint of $\xi=1$. After checking convergence, 22-point grids were used for this transformation. For other element substitutions, where the integrands are very smooth, 6-point equispaced integration grids were used (see the convergence in Fig.\ref{fig3}).


In all TI calculations, a 100-ps NVT-ensemble MD simulation was conducted at each integration grid point. The temperature was maintained at 300 K using the Nose-Hoover thermostat. The mass of hydrogen atoms was set to 2.0 amu, and the MD timestep was set to 1.0 fs. I ensured that the spilling factor for the structures appearing during the calculation remained below 0.02. This indicates that the structures encountered during the calculation are within the interpolation region of the training data.

The TI calculations to correct the results of the MLFFs via Eqs.~(\ref{eq14}) and (\ref{eq15}) were performed using 3-point equispaced grids. At each integration grid point, a 10-ps NVT-ensemble MD simulation at 300 K was conducted. The same FP method used for the training runs was employed.

The local potential gaps $\Delta \phi$ for each solution system were computed from 1s levels of oxygen atoms in water molecules by using the method proposed in the previous study.~\cite{Jinnouchi_npjComputMater_2024, Jinnouchi_arXiv_2024} The computed values of the 1s levels and $\Delta \phi$ are listed in Supplementary Table S2.

\begin{table*}[t]
\caption{Solvation free energies of neutral ion pairs compared with experimental values. The values in the first and second columns for each element are the results from the particle insertion and element substitution methods, respectively. Experimental values are shown in parentheses.$^\mathrm{a}$ The unit is eV.}
\label{table3}
\centering 
\begin{tabular}{p{36mm} p{34mm} p{34mm} p{34mm} p{34mm}}
\hline
              &F$^{-}$                  &Cl$^{-}$                    &Br$^{-}$                     &I$^{-}$                     \\
\hline
H$^{+}$   &$-$15.64$\pm$0.08   &$-$14.36$\pm$0.11      &$-$14.10$\pm$0.09     &$-$13.69$\pm$0.09    \\
             &$-$15.69$\pm$0.12   &$-$14.39$\pm$0.11      &$-$14.18$\pm$0.12     &$-$13.70$\pm$0.11    \\
             &($-$15.89)               &($-$14.60)                   &($-$14.32)                   &($-$13.94)                \\
\hline
Li$^{+}$   &$-$9.72$\pm$0.09   &$-$8.44$\pm$0.10      &$-$8.18$\pm$0.09      &$-$7.78$\pm$0.10    \\
             &$-$9.73$\pm$0.12   &$-$8.42$\pm$0.10      &$-$8.22$\pm$0.11     &$-$7.74$\pm$0.11    \\
             &($-$9.93)               &($-$8.64)                   &($-$8.37)                   &($-$7.98)                \\
\hline
Na$^{+}$   &$-$8.66$\pm$0.09   &$-$7.38$\pm$0.10     &$-$7.12$\pm$0.09      &$-$6.71$\pm$0.10    \\
             &$-$8.67$\pm$0.12   &$-$7.37$\pm$0.10      &$-$7.16$\pm$0.11     &$-$6.68$\pm$0.11    \\
             &($-$8.84)               &($-$7.54)                   &($-$7.25)                   &($-$6.88)                \\
\hline
K$^{+}$   &$-$7.88$\pm$0.10   &$-$6.60$\pm$0.12     &$-$6.33$\pm$0.10      &$-$5.93$\pm$0.11    \\
             &$-$7.90$\pm$0.13   &$-$6.59$\pm$0.12      &$-$6.39$\pm$0.13     &$-$5.91$\pm$0.12    \\
             &($-$8.09)               &($-$6.80)                   &($-$6.52)                   &($-$6.14)                \\
\hline
Rb$^{+}$  &$-$7.62$\pm$0.09   &$-$6.34$\pm$0.11     &$-$6.08$\pm$0.09      &$-$5.67$\pm$0.10    \\
             &$-$7.62$\pm$0.12   &$-$6.32$\pm$0.11      &$-$6.11$\pm$0.12     &$-$5.63$\pm$0.11    \\
             &($-$7.86)               &($-$6.57)                   &($-$6.29)                   &($-$5.90)                \\
\hline
Cs$^{+}$  &$-$7.30$\pm$0.11   &$-$6.02$\pm$0.13     &$-$5.76$\pm$0.12      &$-$5.35$\pm$0.13    \\
             &$-$7.33$\pm$0.12   &$-$6.03$\pm$0.11      &$-$5.82$\pm$0.12     &$-$5.34$\pm$0.11    \\
             &($-$7.54)               &($-$6.24)                   &($-$5.97)                   &($-$5.58)                \\
\hline
\end{tabular}
\raggedright
\footnotesize{\textsf{a From Refs.~\cite{Klots_JPC_1981, Tissandier_JPCA_1998}.}}
\end{table*}

\begin{figure}
\centering
\includegraphics[width=0.5\textwidth,angle=0]{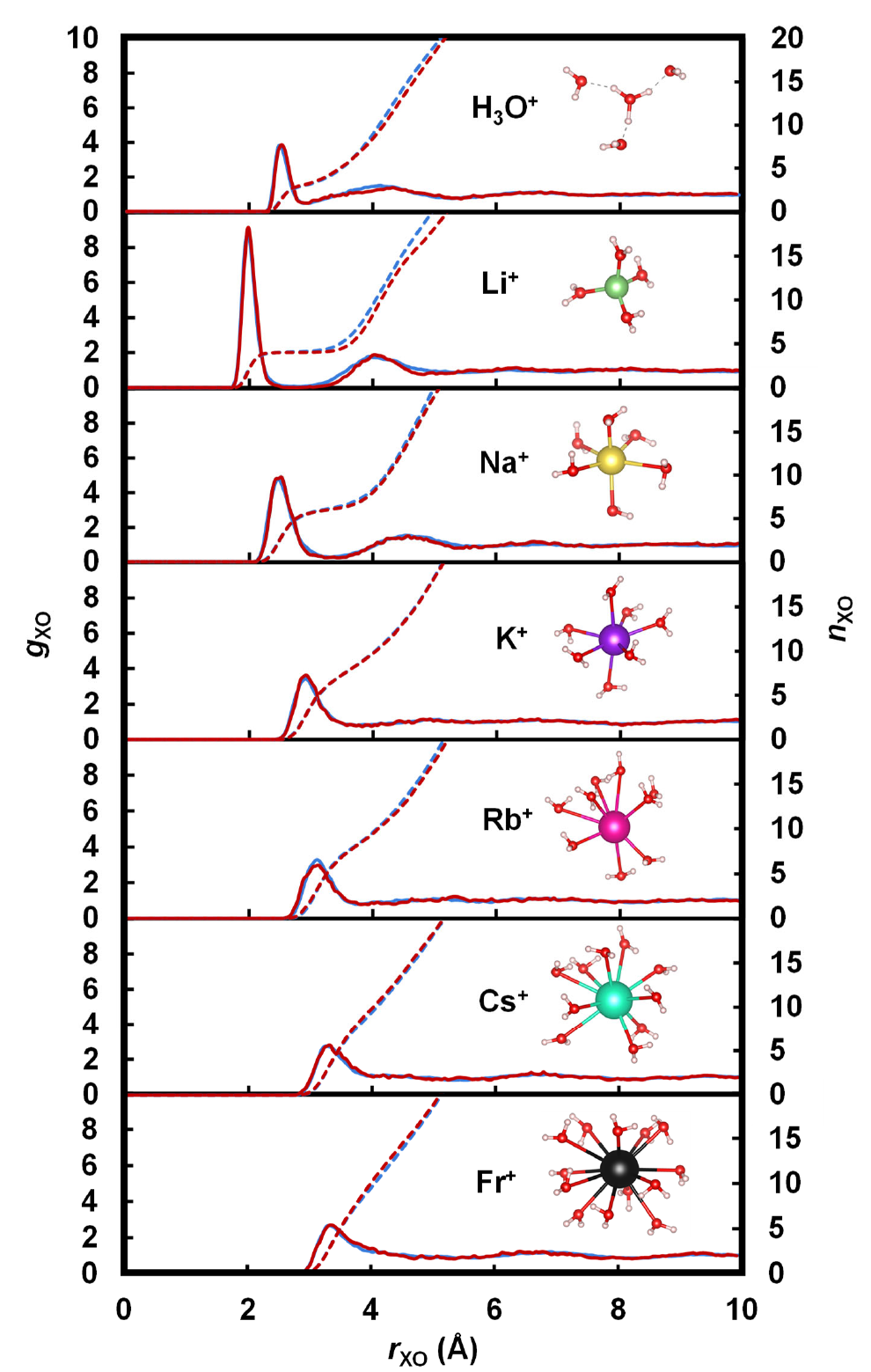}
\caption{RDFs ($g_\mathrm{XO}$) (solid lines) and RINs ($n_\mathrm{XO}$) (dashed lines) between cations (X) and O atoms in water, calculated by RPBE+D3 (red) and MLFFs (blue). For the proton, X represents O in H$_\mathrm{3}$O$^+$.}
\label{fig4}
\end{figure}

\begin{figure}
\centering
\includegraphics[width=0.5\textwidth,angle=0]{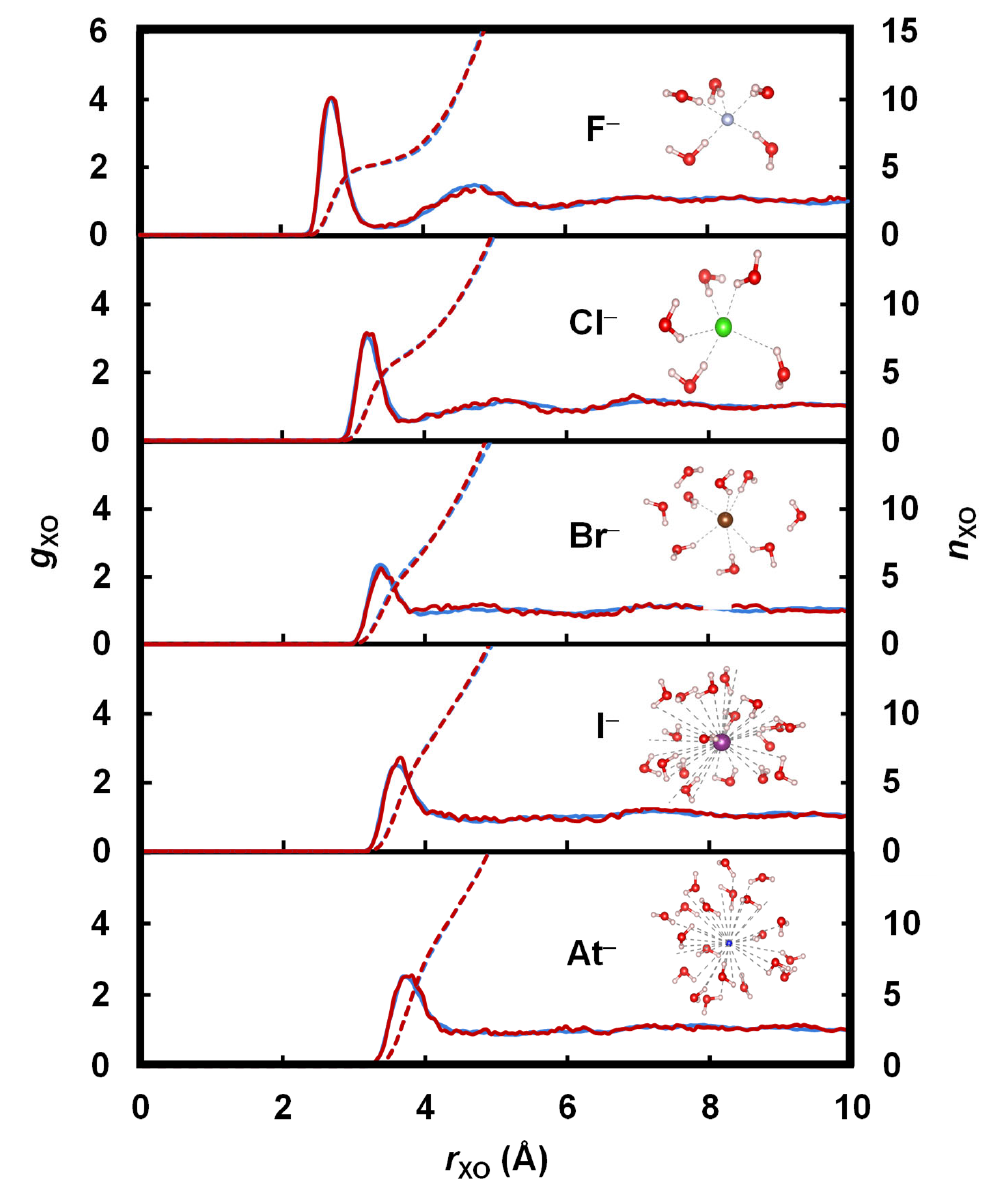}
\caption{RDFs ($g_\mathrm{XO}$) and RIN ($n_\mathrm{XO}$) between anion (X) and O atoms in water.}
\label{fig5}
\end{figure}

\section{\label{section3}Results and discussion}

\subsection{\label{section3_1}Real potentials}

Figures~\ref{fig2} (a) and (b) show the real potentials of the proton (H$^{+}$), alkali metal cations (Li$^{+}$, Na$^{+}$, K$^{+}$, Rb$^{+}$, Cs$^{+}$, and Fr$^{+}$), and halide anions (F$^{-}$, Cl$^{-}$, Br$^{-}$, I$^{-}$, and At$^{-}$) computed using the atom insertion method and the element transformation method. The values are also listed in Supplementary Table~S3. Here, all the results shown in these figures are the FP results after applying the correction to the MLFF results using Eq.~(\ref{eq16}). As mentioned earlier, the difference between the MLFF and FP calculations was very small, with an average difference of 50 meV, which is also worth noting. Despite the completely different coupling pathways, the two methods yield identical real potentials within the statistical error bars. This result indicates good reliability and reproducibility of the computations.

The sum of the real potential for a cation and that for an anion equals the solvation free energy of a neutral ion pair, which can be precisely determined in experiments.~\cite{Klots_JPC_1981, Tissandier_JPCA_1998} The calculated values are compared with the experimental ones in Table~\ref{table3}. While the calculations reproduce the experimental solvation free energies, close comparisons indicate that the RPBE+D3 functional systematically yields weaker ion-water interactions by an average of 0.1 eV per ion.

Although the results of the two TI calculations are identical, the element substitution method leads to a computationally more feasible thermodynamic pathway than particle insertion. As demonstrated by Fig.\ref{fig3} (a) (see also Ref.~\cite{Jinnouchi_arXiv_2024}), in the case of Na$^{+}$ cation insertion, the TI calculation shows a steep change in the integrand near $\lambda_\mathrm{I}=0$, where the not-yet-interacting Na$^{+}$ cation can come very close to other atoms. Additionally, the absolute value of the integrand is relatively large along the entire coupling pathway, and the integrand drops near $\lambda_\mathrm{II}=1$, where the fully interacting Na$^{+}$ cation diffuses, escaping from the restraining model potential.
In contrast, the element substitution from K$^{+}$ to Na$^{+}$ results in a very smooth and small integrand, as shown in Fig.\ref{fig3} (d). Consequently, the integrand converges more quickly with respect to the number of integration grids and the MD simulation time, as illustrated in Fig.\ref{fig3} (e) and (f). It is also noteworthy that the statistical error bars resulting from the element substitution are negligibly small. Therefore, once the real potential of one ion, such as K$^{+}$ in this example, is computed using the particle insertion method, the real potential of another ion, Na$^{+}$, can be easily obtained.

\subsection{\label{section3_2}Solvation structures}

Now, I detail the solvation structures. Radial distribution functions (RDFs) ($g_\mathrm{XO}$) and running integration numbers (RINs) ($n_\mathrm{XO}$) between ions and oxygen atoms in water, computed by the MLFF and FP methods, are shown in Figs.\ref{fig4} and \ref{fig5}. Distances between the ion and the first- and second solvation shells, as well as the number of water molecules in these solvation shells, are listed in Supplementary Tables S4 and S5 for comparison with previous results. For all ions, the MLFFs reproduce the RDFs calculated by the FP method. The first peak in $g_\mathrm{XO}$ for both alkali metal cations and halogen anions shifts outward as the atomic number increases. The peak positions are in good agreement with previous results obtained by MD simulations using conventional force fields and FP methods,~\cite{Dang_JCP_1993, Tuckerman_JCP_1995, Lee_JPC_1996, Toth_JCP_1996, Koneshan_JPCB_1998, Ramaniah_JCP_1999, White_JCP_2000, Raugei_JACS_2001, Lyubartsev_JCP_2001, Hribar_JACS_2002, Grossfield_JACS_2003, Ayala_JCP_2003, Izvekov_JCP_2005, Heuft_F_JCP_2005, Heuft_I_JCP_2005, Lamoureux_JPCB_2006, Dangelo_IC_2010, Bankura_JCP_2013} as well as experimental results.~\cite{Lawrence_JCP_1967, Narten_JPC_1970, Caminiti_JCP_1980, Musinu_JCP_1984, Copestake_JP_1985, Skipper_JP_1989, Neilson_ARPC_1991, Cartailler_JP_1991, Ohtaki_CR_1993, Tanida_JCS_1994, DAngelo_JCP_1994, Enderby_CSR_1995, Fulton_JCP_1996, Beudert_JCP_1997, Wallen_JPCA_1997, Ferlat_PRB_2001, Kameda_BCSJ_2006, Filipponi_PRL_2003, Bowron_JP_2009, Mähler_IC_2012} However, close examination indicates that the RPBE+D3 functional appears to yield slightly larger solvation shells compared to those obtained by experiments and previous simulations using other functionals. This trend is consistent with the weaker ion-water interactions observed in the real potentials.

\begin{figure}
\centering
\includegraphics[width=0.45\textwidth,angle=0]{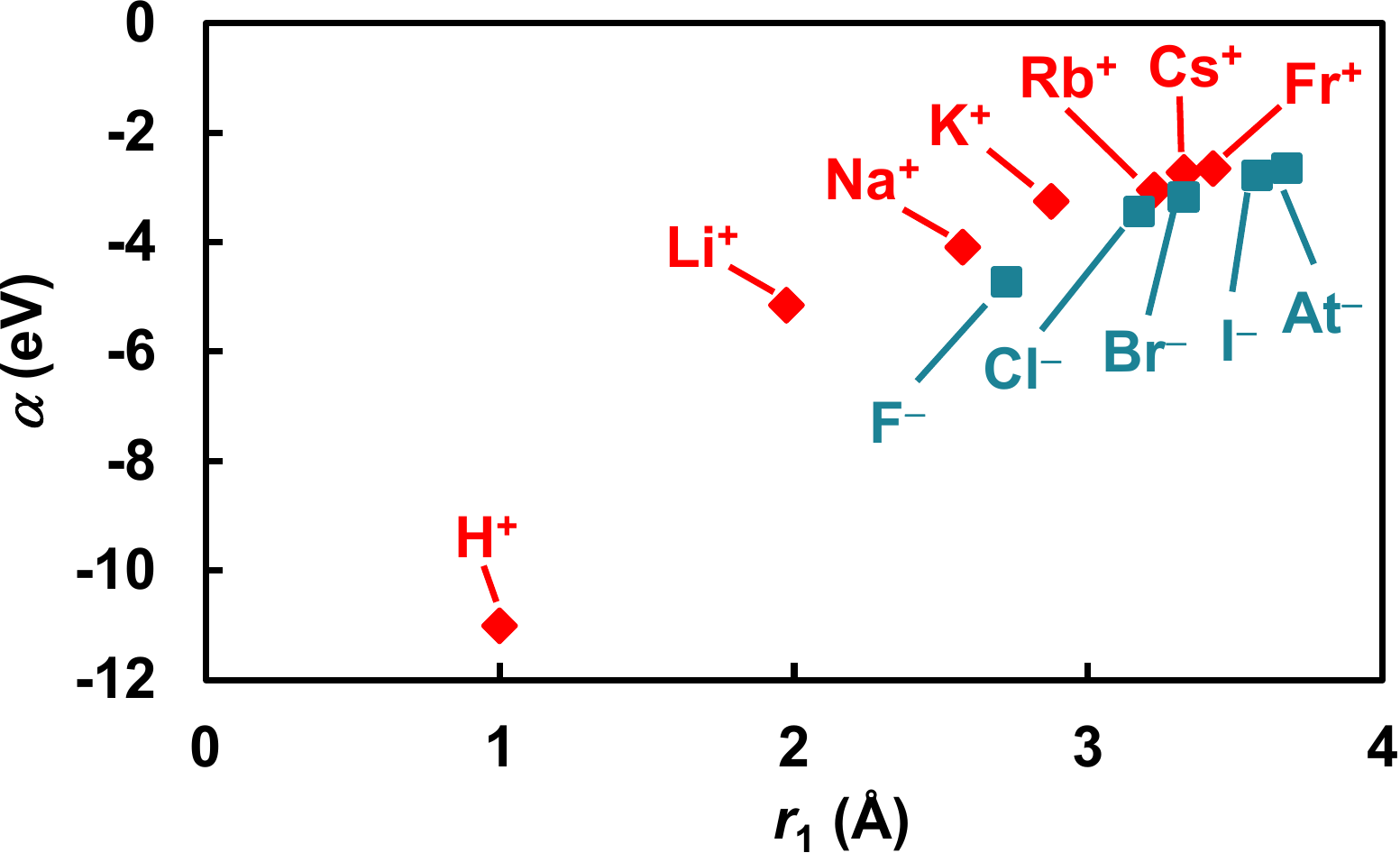}
\caption{Real potential $\alpha$ of a single ion versus the position $r_\mathrm{1}$ of the maximum point of the first peak in the RDF $g_\mathrm{XO}$ between the ion (X) and O atoms in water.}
\label{fig6}
\end{figure}

\begin{figure}
\centering
\includegraphics[width=0.45\textwidth,angle=0]{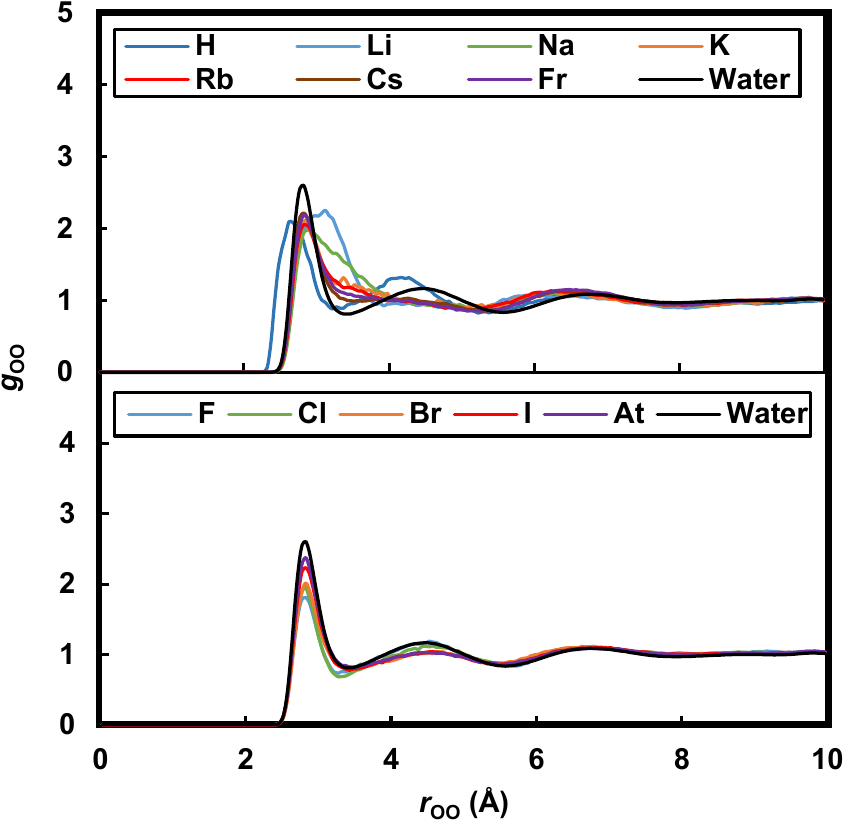}
\caption{RDFs between oxygen atoms of water molecules in the first solvation shell and those of other water molecules in the solutions containing cations (upper panel) and anions (lower panel), compared to the RDF between oxygen atoms in pure water. The first solvation shell is defined as the region up to the first minimum in $g_\mathrm{XO}$.}
\label{fig7}
\end{figure}

The Born model suggests that the solvation free energy is inversely proportional to the ionic radius:
\begin{align}
\Delta G_\mathrm{solv} &= -\frac{1}{2r} \left( 1 - \frac{1}{\epsilon_\mathrm{r}} \right), \label{eq18}
\end{align}
where $r$ is the ionic radius, and $\epsilon_\mathrm{r}$ is the relative permittivity of the solvent. This inverse dependence is observed in the real potentials of both cations and anions versus the position of the first solvation shell, as shown in Fig.~\ref{fig6}. 
However, the simple model cannot explain why the real potential of an anion is lower than that of a cation of similar size, as shown in Fig.~\ref{fig6}. In a previous study using a polarizable force field, Grossfield~\cite{Grossfield_JCP_2004} discovered that the solvation of anions does not cause as much disruption in the surrounding water structure as the solvation of cations does. The same trend is observed in my MD simulations using the MLFFs. 
As shown in Fig.~\ref{fig7}, the RDF between water molecules in the first solvation shell of a cation and other water molecules differs significantly from the RDF of bulk water without ion. Similar to the RDFs reported in previous studies, the height of the first peak diminishes to roughly the same level, and the first minimum disappears, especially for smaller ions. As pointed out by Grossfield, the reduction in the first peak is due to the volume excluded by the ion, and the diminished minimum indicates that water molecules in the first solvation shell have other water molecules packed around them without forming typical hydrogen bonds. The differences from bulk water gradually diminish with increasing ionic radius, indicating that the original water structure is restored as ion-water interactions decrease. In other words, these results suggest there is an energetic trade-off between stabilizing cation-water interactions and stabilizing water-water interactions.
In contrast, the RDFs for halide anions are very similar to those of bulk water. Although the height of the first peak diminishes due to the volume excluded by the ion, the first minimum and second maximum appear at positions identical to those in bulk water. These results indicate that anion solvation does not significantly disrupt the water-hydrogen bonding network.

In summary, the MLFFs accurately replicate the dependence of the real potential of cations and anions on ion size and reproduce the differences in solvation structure between cations and anions, as discovered by a polarizable force field based on physical laws.

\section{\label{section4}Conclusions}

The real potentials of protons, alkali cations, and halide anions were computed using both the particle insertion method and the element substitution method. In both methods, MLFFs trained on the FP training datasets were used to accelerate the MD simulations required for the TI calculations. This scheme allows for the efficient computation of most of the free energy changes, which would typically require long simulation times ranging from several nanoseconds to tens of nanoseconds, by using the MLFF surrogate models in the TI calculations. The residual errors of the MLFFs are then precisely corrected by TI calculations over tens of picoseconds, transitioning from the MLFF potential energies to those of the FP method. Both the particle insertion and element substitution methods yielded highly reversible pathways between the two endpoints, and these two entirely different thermodynamic pathways resulted in identical real potentials within statistical error bars. This confirms that both methods can provide reliable and reproducible free energy changes. Particularly, the element substitution method achieves smooth thermodynamic pathways, enabling efficient and accurate TI calculations.

Although both thermodynamic integration methods yield consistent real potentials, it was found that the RPBE+D3 functional used in this study provides slightly weaker ion-water interactions, approximately 0.1 eV per ion, compared to experimental values. However, the RPBE+D3 functional reasonably reproduces the inverse proportionality of real potential to ion size and the easier solvation of anions compared to cations, as revealed by experiments and calculations using physics-based polarizable force fields. The two TI methods proposed in this study are versatile and can accurately compute free energy changes for a wide range of phenomena, such as the insertion of molecules or ions into liquids and the doping of elements into solids. They also provide a powerful means to verify the correctness of these calculations against each other.

\section*{Acknowledgement}
In this study, I received valuable advice on computational techniques and the use of software from Professor Georg Kresse and Dr. Ferenc Karsai. We would like to express our deep gratitude for their guidance.

\section*{Code availability}

The VASP code is distributed by the VASP Software GmbH. The machine learning modules will be included in the release of vasp.6.3.

\section*{Data availability}

The data that support the findings of this study are available from the corresponding author upon reasonable request.

\bibliographystyle{unsrt}
\bibliography{msNotes}

\end{document}